\documentclass[namedreferences]{solarphysics}

\usepackage[optionalrh]{spr-sola-addons} 

\usepackage[utf8]{inputenc}
\usepackage[T1]{fontenc}
\usepackage[english]{babel}

\usepackage{graphicx}
\usepackage[caption = false]{subfig}

\usepackage{tabularx}
\usepackage{enumitem}
\usepackage{multirow}
\usepackage{makecell}
\usepackage{booktabs,dcolumn,eurosym}
\usepackage{hhline}
\usepackage{rotating}
\usepackage{pdflscape}

\usepackage{mathtools}
\usepackage[title]{appendix}
\usepackage{academicons}
\usepackage{xcolor}
\usepackage{xurl}
\usepackage{hyperref}

\renewcommand{\orcid}[1]{\href{https://orcid.org/#1}{\textcolor[HTML]{A6CE39}{\includegraphics[]{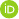}}}}

\begin{document}

\begin{article}
\begin{opening}

\title{The high-energy protons of the ground level enhancement (GLE74) event on 11 May 2024}

\author{A.~\surname{Papaioannou\orcid{0000-0002-9479-8644}}$^1$}
\author{A. ~\surname{Mishev\orcid{0000-0002-7184-9664}}$^2$}
\author{I. ~\surname{Usoskin\orcid{0000-0001-8227-9081}}$^2$}
\author{B.~\surname{Heber\orcid{0000-0003-0960-5658}}$^{3}$}
\author{R.~\surname{Vainio\orcid{0000-0002-3298-2067}}$^4$}
\author{N.~\surname{Larsen\orcid{0000-0003-4713-1350}}$^2$}
\author{M.~\surname{Jarry\orcid{0000-0002-3653-1722}}$^1$}
\author{A.P.~\surname{Rouillard\orcid{0000-0003-4039-5767}}$^5$}
\author{N.~\surname{Talebpour Sheshvan\orcid{0000-0002-9774-9047}}$^5$}
\author{M.~\surname{Laurenza\orcid{0000-0001-5481-4534}}$^6$}
\author{M.~\surname{Dumbovi{\'c} \orcid{0000-0002-8680-8267}}$^7$}
\author{G.~\surname{Vasalos\orcid{0000-0003-0490-2717}}$^1$}
\author{J.~\surname{Gieseler\orcid{0000-0003-1848-7067}}$^4$}
\author{S.~\surname{Koldobskiy\orcid{0000-0001-9187-0383}}$^2$}
\author{O.~\surname{Raukunen\orcid{0000-0001-8346-5281}}$^8$}
\author{C.~\surname{Palmroos\orcid{0000-0002-7778-5454}}$^4$}
\author{M.~\surname{H{\"o}rl{\"o}ck\orcid{0009-0006-2565-5670}}$^3$}
\author{M.~\surname{K{\"o}berle\orcid{0000-0001-9751-8137}}$^3$}
\author{R. F.~\surname{Wimmer-Schweingruber\orcid{0000-0002-7388-173X}}$^3$}
\author{A.~\surname{Anastasiadis\orcid{0000-0002-5162-8821}}$^1$}
\author{P.~\surname{K\"{u}hl \orcid{0000-0002-3758-9272}}$^3$}
\author{E.~\surname{Lavasa\orcid{0000-0003-1192-0868}}$^{1}$}

\runningauthor{A. Papaioannou \textit{et al.}}

\runningtitle{GLE74: First Observations}

\institute{$^1$ Institute for Astronomy, Astrophysics, Space Applications and Remote Sensing (IAASARS), National Observatory of Athens, I. Metaxa \& Vas. Pavlou St., 15236 Penteli, Greece}
\institute{$^2$ Space Physics and Astronomy Research Unit and Sodankyl\"a Geophysical Observatory, University of Oulu, Oulu, Finland}
\institute{$^3$  Institut f\"ur Experimentelle und Angewandte Physik, Christian-Albrechts-Universit\"at zu Kiel, 24118 Kiel, Germany }
\institute{$^4$ Department of Physics and Astronomy, University of Turku, 20500 Turku, Finland}
\institute{$^5$  IRAP, Université Toulouse III—Paul Sabatier, CNRS, CNES, Toulouse, France }
\institute{$^6$ INAF-Istituto di Astrofisica e Planetologia Spaziali, Via del Fosso del Cavaliere, 100, I-00133 Roma, Italy}
\institute{$^7$ University of Zagreb, Faculty of Geodesy, Hvar Observatory, Croatia}
\institute{$^8$ Aboa Space Research Oy (ASRO), Tierankatu 4B, 20520 Turku, Finland}

\begin{abstract}
High energy solar protons were observed by particle detectors aboard spacecraft in near-Earth orbit on May 11, 2024 and produced the 74$^\mathrm{th}$ ground level enhancement (GLE74) event registered by ground-based neutron monitors. This study involves a detailed reconstruction of the neutron monitor response, along with the identification of the solar eruption responsible for the emission of the primary particles, utilizing both in situ and remote-sensing. Observations spanning proton energies from a few MeV to around 1.64 GeV, collected from the Solar and Heliospheric Observatory (SOHO), the Geostationary Operational Environmental Satellite (GOES), the Solar Terrestrial Relations Observatory (STEREO-A), and neutron monitors, were integrated with records of the associated solar soft X-ray flare, coronal mass ejection, and radio bursts, to identify the solar origin of the GLE74. Additionally, a time-shift analysis was conducted to link the detected particles to their solar sources. Finally, a comparison of GLE74 to previous ones is carried out. GLE74 reached a maximum particle rigidity of at least 2.4 GV and was associated with a series of type III, type II, and type IV radio bursts. The release time of the primary solar energetic particles (SEPs) with an energy of 500 MeV was estimated to be around 01:21 UT. A significant SEP flux was observed from the anti-Sun direction with a relatively broad angular distribution, rather than a narrow, beam-like pattern, particularly during the main phase at the particle peak flux. Comparisons with previous GLEs suggest that GLE74 was a typical event in terms of solar eruption dynamics.
\end{abstract}

\end{opening}
\section{Introduction}
\label{sec:intro}

Ground-level enhancements (GLEs) are produced in the Earth atmosphere by relativistic solar energetic particle (SEPs), requiring acceleration mechanisms capable of producing particles with rigidities of $\geq$1 GV \citep{macau_mods_00083703}. These particles must have sufficient energy and flux to initiate an atmospheric cascade detectable by neutron monitors (NMs) on the ground \citep[e.g.,][]{Poluianov2017}. Relativistic protons are particularly valuable, due to their rapid propagation, for identifying SEP sources at the Sun \citep{aschwanden2012gev}. SEPs are associated with both intense solar flares and fast, wide coronal mass ejections (CMEs). Therefore, pinpointing their precise acceleration site remains challenging. Detailed investigations of individual GLEs have been conducted \citep[e.g.,][]{bombardieri2008improved, Kleinetal2014,2022A&A...660L...5P, Kleinetal2022}, however the exact conditions and processes leading to such powerful SEP events are not yet fully understood.

GLEs are rare phenomena, occurring at an average rate of ~0.9 events per year \citep[i.e., 76 events over $\sim$83 years;][]{vainio2017solar, macau_mods_00083703}. These events have primarily been detected by ground-based NMs, with their lower-energy components observed by spacecraft in near-Earth space \citep[e.g.][]{2018SoPh..293..136M, 2017SoPh..292...10K,kocharov2021multiple, 2022A&A...660L...5P, Kocharov2023, 2023SpWea..2103191M, 2024A&A...682A.106K}. 
However, the analysis of GLEs has been limited by the poor coverage of high-energy (E$>$200 MeV) proton observations recorded on board spacecraft that could shed light into the spectral characteristics of these events. \citet{2015A&A...576A.120K} emphasized that the reliance on NM data alone introduces uncertainties in determining the high-energy tail of SEP spectra, as NM responses depend on complex atmospheric and geomagnetic filtering effects. These authors further demonstrated that such detectors as the Electron Proton Helium Instrument \citep[EPHIN;][]{1995SoPh..162..483M} are capable of producing data from 250 MeV up to 1.6 GeV \citep{2016SoPh..291..965K} and, in particular, measure SEP protons between 100 MeV and above 800 MeV \citep{2015A&A...576A.120K}. This energy region is the focus of the Horizon Europe SPEARHEAD project\footnote{\url{https://spearhead-he.eu/}}.

GLE74 occurred on 11 May 2024 during significantly disturbed magnetospheric conditions. The latter resulted from a series of Earth-directed CMEs unleashed from the Sun from 08~May onwards, impacting our planet and leading to a remarkable G5-class ``extreme'' geomagnetic storm\footnote{\url{https://www.swpc.noaa.gov/noaa-scales-explanation}} that lasted from 10--13 May 2024. The storm peaked on 11~May 2024 leading to a Disturbance Storm Time (Dst) index of --412 nT, marking it as one of the most intense geomagnetic events since 1957 \citep[see details in][]{2025ApJ...979...49H}. In terms of cosmic ray particles, a large Forbush Decrease (FD) was recorded globally by many NMs across Earth. However, the FD had different impacts depending on the location of the observing NM. In particular, on 10 May 2024 at $\sim$17:00~UT high-latitude (polar) NMs (i.e. South Pole; SOPO), detected a sharp reduction in cosmic ray counts that led to an FD with a magnitude of $\sim$18\% \footnote{\url{https://www.nmdb.eu}}.  
Meanwhile, for NM stations at mid and low latitudes, FDs with magnitudes of $\sim$ 10\% (Lomnicky-Stit; LMKS, Rome; ROME) were observed, respectively. Moreover, for these stations, the geomagnetic storm played a dominant role by temporarily altering Earth's magnetosphere and influencing how cosmic rays reach the surface leading to the recording of enhancements on 11 May 2024 possibly due to ``magnetospheric" responses \citep[see details in][]{2005JGRA..110.9S20B}. Consequently, the recorded increase in the counting rate of NMs during GLE74 reflected a combination of several competing factors in addition to the SEP-enhanced flux: the gradual recovery from the FD and the geomagnetic storm’s effects on cosmic ray entry to the inner magnetosphere. 

This work combines ground-based and near-Earth particle observations of GLE74 with data on CME evolution, contextual solar activity, and SEP flux modeling derived from NM recordings.



\section{Overview of GLE74} \label{subsec:overview}
\begin{figure}[h!]
\centering
\includegraphics[width=\columnwidth]{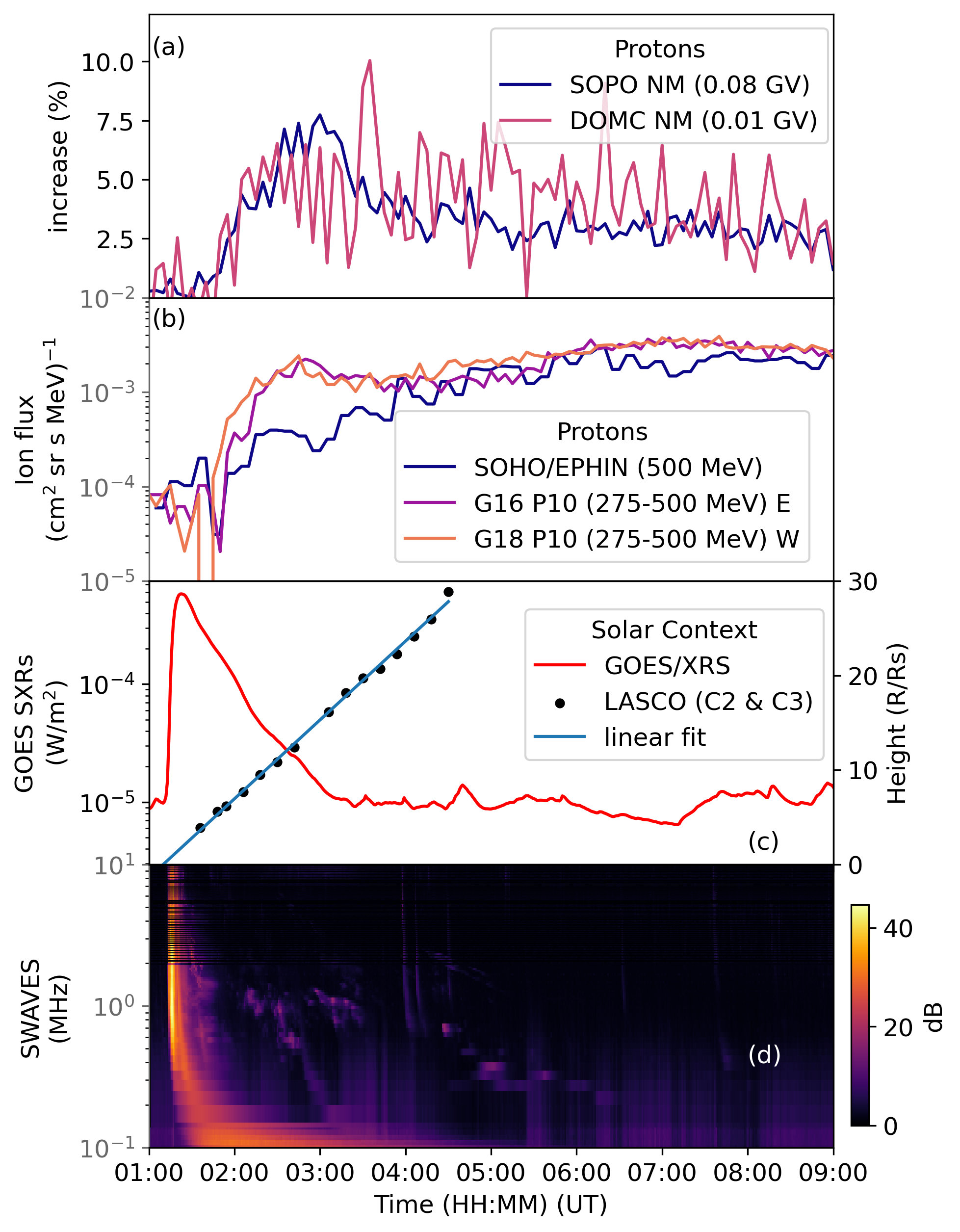}
\caption{GLE74 on 11 May 2024. Panel (a): Count rate increase (in percent) of SOPO and DOMC NMs based on 5-minute de-trended NM data. Panel (b): SOHO/EPHIN and GOES/SEISS proton flux. Panel (c): SXR flux observed by GOES, denoting an X5.8 solar flare (red curve; left axis). The height time of the CME evolution is shown with the black circles from measurements at the plane-of-sky near the CME leading edge (taken by the LASCO CME CDAW catalog). The solid blue line is a linear fit to the height and extrapolated back to the surface of the Sun. Panel (d): Dynamic radio spectrum observed by STEREO-A/WAVES (SWAVES).} 
\label{fig:fig1}
\end{figure}

GLE74 was detected by multiple NMs worldwide on 11 May 2024 (see e.g. Table \ref{tab:table1}). Figure~\ref{fig:fig1} provides an overview of selected observations. Among the standard-design NM stations, the peak count rate increase was highest on the Antarctic plateau, reaching approximately 10.0\% at Dome C (DOMC) and 7.7\% at the South Pole (SOPO). Lead-free (bare) NMs at the same locations recorded stronger responses, with 16.0\% at Dome B (DOMB) and 8.9\% at South Pole Bare (SOPB) (not shown). Energetic protons were also observed by the Solar and Heliospheric Observatory (SOHO)/EPHIN at $E$=500 MeV and the Geostationary Operational Environmental Satellite (GOES)/Space Environment In-Situ Suite \citep[SEISS;][]{kress2020goes} in the 275--500 MeV energy range. Figure \ref{fig:fig1} (b) illustrates the highest SOHO/EPHIN proton channel ($E$=500 MeV) alongside the P10 channel (275--500 MeV) from the east (GOES16; purple line) and the west (GOES18; pink line) GOES/SEISS measurements. GLE74 was associated with a powerful X5.8-class solar flare, which began at 01:12 UT and peaked at 01:23 UT (see Fig.~\ref{fig:fig1} (c); red solid line), as well as a halo CME traveling at 1614 km/s, first observed by LASCO at 01:36~UT (see Fig.~\ref{fig:fig1} (c); height time points from LASCO\footnote{\url{https://cdaw.gsfc.nasa.gov/CME\_list/UNIVERSAL\_ver2/2024\_05/yht/20240511.013605.w360h.v1614.p304g.yht}}). The source of the active region appears to be NOAA AR13668\footnote{\url{https://solarmonitor.org/?date=20240511}}, located near the western limb at S15W55 from Earth's perspective. However, AR13664 (S17W62) was seemingly merging with AR13668 and thus the responsible AR was 13664/8 \citep[see details in][]{2024A&A...692A.112W}. In addition, from metric to kilometric wavelengths (radio domain), type III, type II, and IV radio bursts were also observed in association with these solar events (see Fig.~\ref{fig:fig1} (d) and details below). 

\section{Neutron monitor data}
\label{sec:neutron}
\begin{figure}[h!]
\centering
\includegraphics[width=\columnwidth]{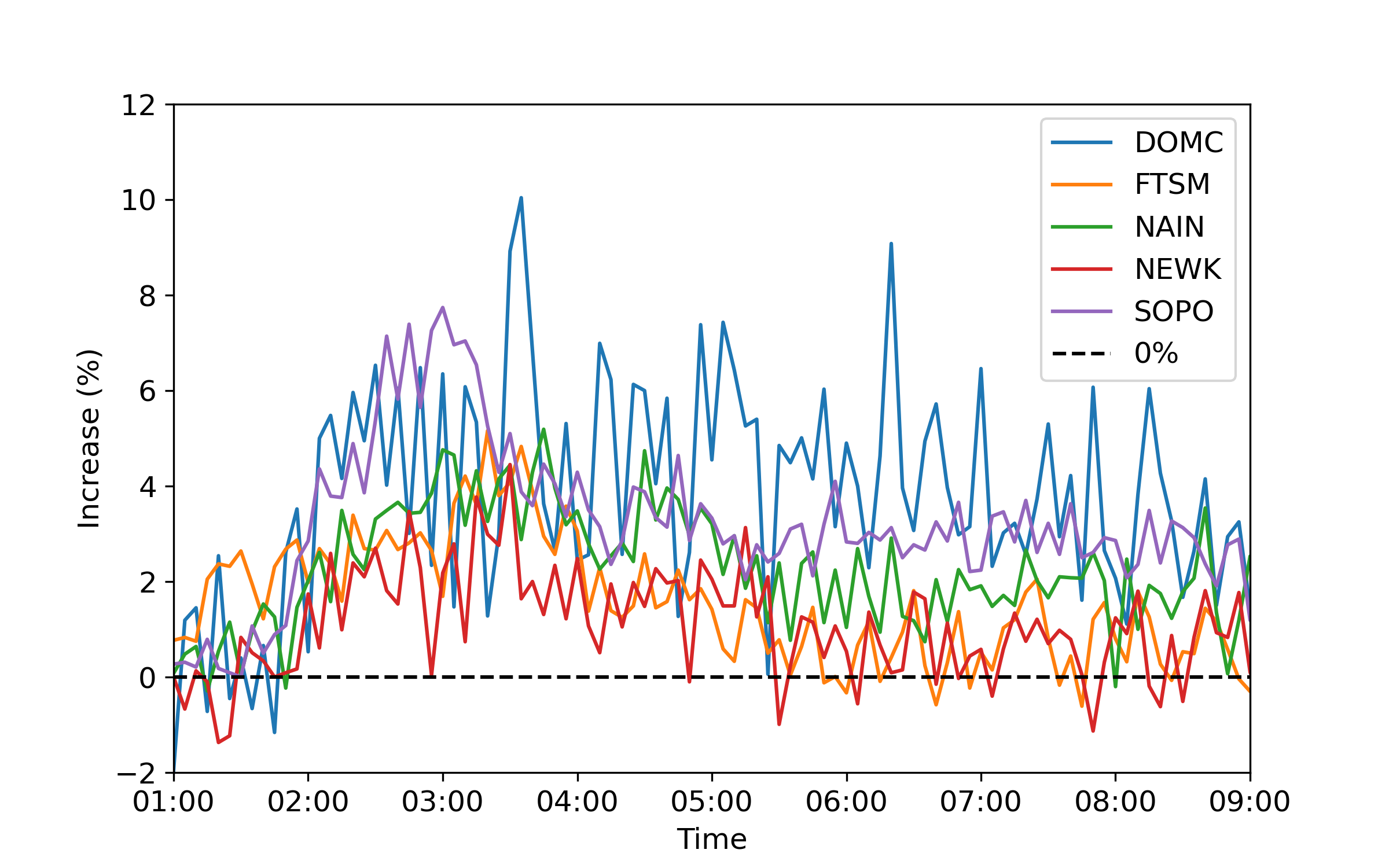}
\caption{Count rate variation of selected NMs during GLE 74. The vertical dashed black line depicts the 0\% level. De-trended 5-min averaged data are used.} 
\label{fig:NM}
\end{figure}

GLEs are identified as relative increases in count rates of different NMs over the background caused by GCRs \citep{vashenyuk2006some, bombardieri2008improved, But09, Mishev16SF}. During GLE74, differences in the time profiles of the cosmic-ray intensity are evident, as presented in Fig.~\ref{fig:NM}. Herein, we use five-minute de-trended NM data \citep{usoskin2020revised} retrieved from the International GLE Database (IGLED)\footnote{\url{https://gle.oulu.fi/}}. One can see that the event exhibited a typical gradual increase and notable anisotropy (see details in Sects. \ref{sec:NMres} and \ref{sec:NMcom}) during the onset since a moderate count-rate increase is recorded by the stations looking in the sunward direction (i.e., FSMT and SOPO). The NMs situated at high-altitude polar stations (i.e., DOMC and SOPO) recorded the largest count-rate increases (see Fig.~\ref{fig:fig1}). The rapid rise as shown by the FSMT and SOPO NM intensity time profiles (Fig.~\ref{fig:NM}) indicates that energetic protons had reasonably good access to the Sun-Earth-connecting field lines. For seventeen NMs and the two bare NMs, the onset and peak times as well as the maximum increases (in percent) were calculated. All results are presented in Table~\ref{tab:table1}.

Inspection of the NM data from various stations around the world (Fig. \ref{fig:NM}) indicated the presence of particles with rigidity of up to $\sim$2 GV. The Newark NM (NEWK), with a vertical nominal cutoff rigidity of 2.4 GV, recorded an increase of marginal significance (4.45\%) that may (or may not) be related to GLE74. Other NM stations at higher cutoff rigidities like i.e. Almaty (AATB; 5.9 GV) and Baksan (BKSN; 5.7 GV) recorded increases of $\sim$2\%. The identification of the exact arrival times and the amplitudes of count rate increases of solar particles at different NMs is challenging, as noted above, due to the fact that solar particles: (a) propagated under very disturbed conditions due to the sequence of several CMEs; (b) a large FD was in progress; (c) a severe geomagnetic storm was at play \citep[details given in ][]{2025ApJ...979...49H}. Nonetheless, solar particles of at least 2.4 GV seem to have arrived at Earth.  

\begin{figure}[h!]
\centering
\includegraphics[width=\columnwidth]{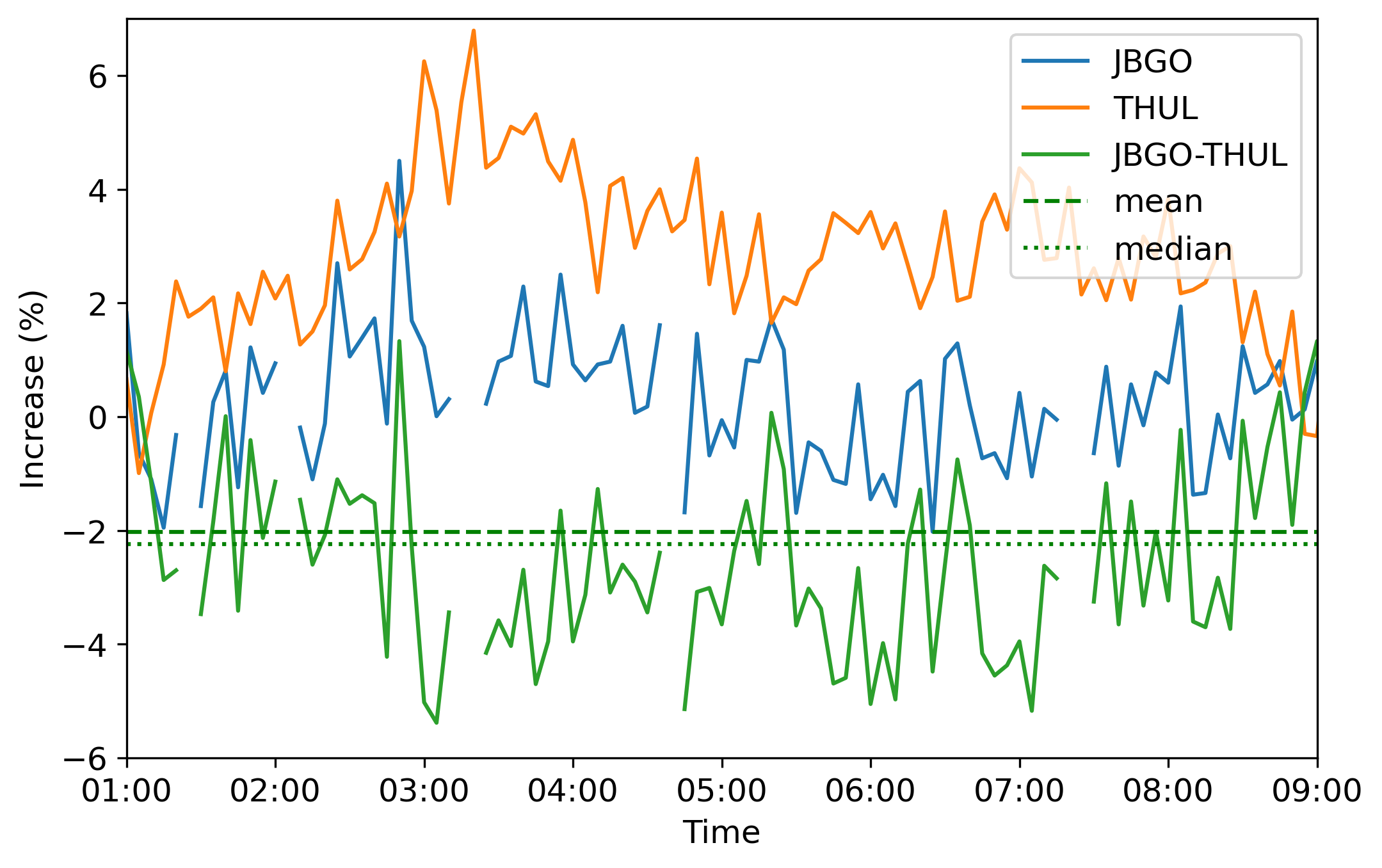}
\caption{Relative variations of the count rates of two sub-polar stations, THUL (north, orange line) and JBGO (south, blue line), and their corresponding difference (green line) are shown. The viewing directions of the two stations approximately represent the sunward (JBGO) and antisunward
(THUL) Parker spiral direction. The horizontal dashed (dotted) line depicts the mean (median) of their corresponding difference.} 
\label{fig:aniso}
\end{figure}

A qualitative study of anisotropy is based on comparison of the count rate variation of (mostly polar) NMs. Figure \ref{fig:aniso} presents an illustration of the count rates of two high latitude NMs - Thule (THUL, 75.6$^{\circ}$ N) and Jang Bogo (JBGO, 74.6$^{\circ}$ S) — which share similar characteristics, with a vertical cutoff rigidity of 0.0 GV and site altitudes of 260 m and 30 m, respectively. As shown in Fig. \ref{fig:aniso}, the difference (green line) fluctuated around 2\% during GLE74, with an absolute value for the mean of 2.02\% and for the median of 2.23\%. We note, that JBGO NM is viewing particles arriving from a direction close to the sunward direction (for details see Fig. \ref{fig:Cones}), whilst  THUL NM is viewing particles arriving from the anti-Sun direction. Because the two NMs have similar energy responses, the different traces in Fig.~\ref{fig:aniso} result from the anisotropy of the incoming solar particles. We emphasize that the difference in count rate variation of the bulk of polar station is in the same order, which implies broad angular distribution of the incoming SEPs, as confirmed by the full i.e. quantitative analysis (see Section \ref{sec:NMres}).

\begin{figure}[h!]
\centering
\includegraphics[width=\columnwidth]{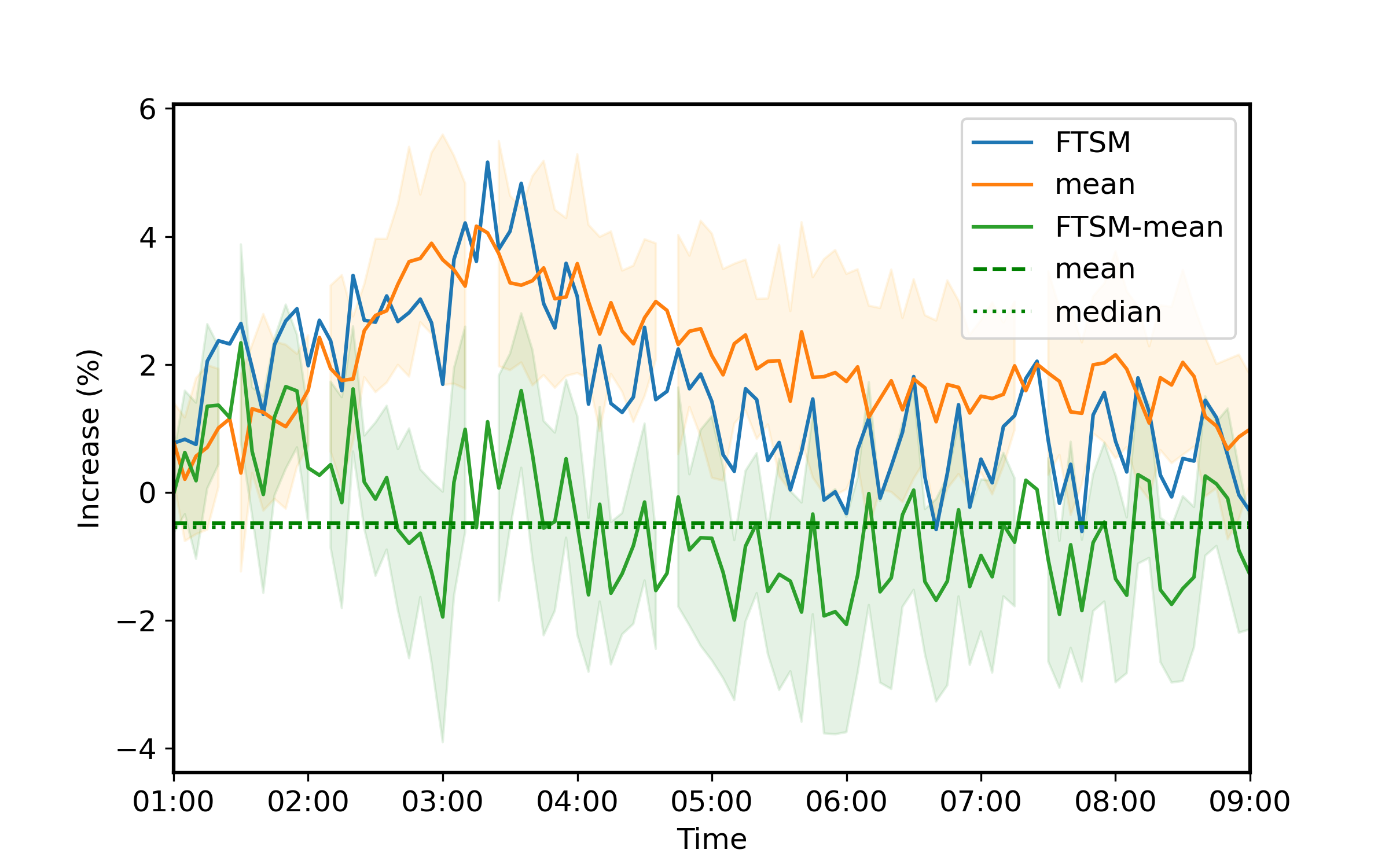}
\caption{Evaluation of high-latitude NMs during GLE74. Variations using the FSMT NM (solid blue line) and the mean of ten high-latitude NMs (solid orange line) and their corresponding difference (green line) are shown. The orange and the green ribbons depict the 1$\sigma$ error.} 
\label{fig:figaniso}
\end{figure}

At this point we used 11 NM stations with a nominal cutoff rigidity $R_{C}<$1.4 GV \citep{kurt2019onset}: Apatity (APTY), Fort Smith (FSMT), Inuvik (INVK), Jang Bongo (JBGO), Mawson (MWSN), Nain (NAIN), Oulu (OULU), Peawanuck (PWNK), Terre Adelie (TERA), Thule (THUL), and Tixie Bay (TXBY) for which de-trended NM data were available. Figure \ref{fig:figaniso} shows a comparison of the averaged data of ten of these stations (orange line) against the recordings of the FSMT NM (blue line). The difference (green line) is small, fluctuating around an absolute value of 0.5\% (mean = 0.48\%, median = 0.54\%). The largest difference was noticed from 02:00--03:00 UT and reached $\sim$2\% during GLE74. 

Table \ref{tab:table1} presents the characteristics of GLE74, including its onset, peak time, and maximum NM count rate increase in percentage. Column 1 lists the names of the neutron monitors (NMs) used in this part of the analysis, identified by their conventional acronyms. Column 2 shows the calculated cut-off rigidity for each NM station (in GV), Column 3 indicates the GLE onset time (in UT) \citep{2022FrASS...973578P}, Column 4 provides the peak time (also in UT), and Column 5 shows the maximum percentage increase recorded at each NM station. These values were derived from 5-minute de-trended NM data \citep{usoskin2020revised}. While a finer time resolution (e.g., 1-minute) could, in principle, offer a better correlation with the solar source, the statistical fluctuations would be excessively large. In addition, the NM count rate increases were complicated by the interplay between the actual signal due to SEPs, the recovery of the FD, and the reduction of the cut-off rigidities of the stations; the latter specifically important for low- and mid-latitude stations. As a result, onset times reflect this uncertainty. 

The high-altitude, high-latitude stations DOMC and SOPO are more sensitive than most NMs as they can detect lower-energy particles \citep{kuwabara2006real,Mishev2021AP}. Consequently, these stations recorded the highest flux intensity during GLE74. Additionally, the bare neutron monitors at these locations (DOMB and SOPB) captured the most pronounced signals of solar particles for this event (see Table \ref{tab:table1}). As expected, bare NMs are relatively more sensitive to low-energy primaries and thus record a higher percentage increase than the standard NM, owing to the soft spectrum of solar cosmic rays \citep{2002ApJ...567..622B}.

\begin{table}[h!]
\centering
\caption[]{Characteristics of GLE74 as recorded by NMs. }
\label{tab:table1}
\begin{tabular}{l|cccc}
\hline 
\hline
Neutron & $R_{c}$ & Onset & Peak & Maximum\\
 Monitor & (GV) & Time (UT)& Time (UT) & Increase (\%)\\
\hline
DOMB  & 0.0 & 01:55$\ast$ & 04:20 & 16.0 \\
SOPB  & 0.5 & 01:45 & 02:45 & 8.92 \\
\hline
APTY  & 0.103 & 01:35 & 04:00 & 6.88 \\
CALG  & 0.41 & 02:10 & 03:25 & 4.71 \\
DOMC  & 0.0 & 02:00 & 03:35 & 10.00\\
FSMT  & 0.178& 01:15 &  03:20 & 5.16 \\
INVK  & 0.0 & 01:30 &  03:20 & 5.03\\    
JBGO  &  0.0& 02:20$\ast$ & 02:50 & 4.50 \\
MRNY  &  0.0& 01:35 & 02:20 & 5.10 \\
MWSN  &  0.196& 02:20$\ast$ & 02:40  & 4.60 \\
NAIN  &  0.24& 02:30$\ast$ & 03:45 & 5.19 \\
NEWK  &  1.021 & 01:30$\ast$ & 03:30 & 4.45\\
OULU  & 0.105 & 01:35$\ast$ & 03:25 & 5.81 \\
PWNK  & 0.278 & 02:30& 03:00 & 5.84 \\
SOPO  & 0.5 & 01:35& 03:00 & 7.74 \\
TERA  & 0.0 & & 04:30 & 2.88 \\
THUL  & 0.0 & 02:25& 03:20 & 6.79 \\
TXBY  & 0.0 & 02:00& 02:55 & 5.54 \\
YKTK  & 0.684 & 02:20& 02:50 & 4.29 \\
\hline
 \multicolumn{5}{p{0.90\columnwidth}}{\textbf{Notes}. The de-trended NM data from the IGLED are used. The top two rows refer to the bare NMs and the rest to the conventional NMs. When onset/peak times could not be reliably identified the relevant entry remained blank. $R_{c}$ are computed specifically for the event, for details see subsection \ref{sec:NMasym}  }\\
\multicolumn{4}{l}{$\ast$ ambiguous due to data fluctuations}
\end{tabular}%
\end{table}

\section{Modelling the neutron monitor response}
\label{sec:NMres}
The spectra and anisotropy of the SEPs leading to a GLE can be unfolded through the modeling of the global NM response and subsequent optimization procedure. Herein, we use a similar approach to \citet{Cramp97}, used by \citet{Bom06, Vas06}. The exact method employed here is described in \citet{Mishev2024SF}.  

The method involves computation of geomagnetic cutoff rigidities and asymptotic directions of all NM stations used in the analysis and least-squares optimization of the difference between modelled and experimental data, that is the simulated over the recorded NM responses. We note that the model must reproduce the response of the stations with statistically significant increases in count rate as well as stations with marginal or zero responses \citep{Cramp97, Mishev2024SF}. The zero responses are particularly important, because constrain the flux and hardness of the SEP spectra, the anisotropy and the apparent source axis position. Here, we conservatively assume all nonpolar stations as zero response, since their count rate increases are due to geomagnetospheric effects and the recovery of the FD, similarly to \citet{Larsen2025}.

During optimization, the initial guess of the model parameters that describe the SEP spectra and angular distribution is selected following a plausible set obtained using analysis of a large number of GLEs \citep{Kocharov2023, Larsen2024all}. For global NM response modelling, we use a new-generation yield function \citep{MishevNMYFJGR2020}, which is in very good agreement with latitude surveys and was recommended for GLE analysis \citep{Nuntiyakul20187181, Xaplanteris2021, Caballero-Lopez20222602}. For the optimization, we employ a combination of the Levenberg-Marquardt algorithm \citep{Lev44,Mar63} and Ridge regression \citep{Tikhonov1995, Huber2019}, leading to robust convergence  \citep{Den83, Engl96, Mishev20057016}. 

\subsection{Calculation of the asymptotic directions}
\label{sec:NMasym}
For the analysis of GLE74, it is necessary first to accurately model the magnetospheric conditions. Herein, we employed the recently developed OTSO tool for the computation of CR trajectories within a realistic model of the Earth's magnetosphere \citep[for details see][]{Larsen_2023_OTSO}. The tool uses a combination of models for the internal magnetic field, created by the Earth's dynamo,  and the external magnetic field, related to the system of magnetospheric currents. For the former, we used the 13th generation of the International Geomagnetic Reference Field \citep[IGRF,][]{Alken2021}, while for the latter the storm-time variant of the Tsyganenko 01 (TSY01S) model \citep{TSY01S} was employed. TSY01S is similar to the earlier Tsyganenko 01 (TSY01) model \citep{TSY01a, TSY01b}, but is parameterized using data only from geomagnetic storm events, providing a better representation of stormy magnetospheric conditions. We emphasize that under disturbed geomagnetospheric conditions, more recent Tsyganenko models, such as TSY01, are recommended to represent the magnetosphere \citep{Kudela04}, moreover, they provide a better overall description of the external magnetosphere. 

For the TSY01S model, it is necessary to provide several input parameters: solar wind speed ($v$), $y$ and $z$ components of the interplanetary magnetic field (IMF$_y$ and IMF$_z$), solar wind dynamic pressure ($P_\mathrm{dyn}$), Dst index, G2 and G3, the latter are variables unique to TSY01S computed using geomagnetic conditions during the hour preceding the event \citep{TSY01b,TSY01S}. The input TSY01S parameters for GLE74 were: $v$\,=\,733.27\,km/s, IMF$_y$\,=\,--1.94\,nT, IMF$_z$\,=\,--25.03\,nT, $P_\mathrm{dyn}$\,=\,23.6\,nPa, Dst\,=\,--397\,nT, G2\,=\,91.8, and G3\,=\,241.2. The solar wind measurements data were taken from both the ACE (\url{https://izw1.caltech.edu/ACE/ASC/}) and SOHO (\url{https://space.umd.edu/pm/crn/}) spacecraft; the Dst index was taken from the World Data Center for Geomagnetism, Kyoto (\url{https://wdc.kugi.kyoto-u.ac.jp/}).

\begin{figure}[h!]
\centering
\includegraphics[width=\columnwidth]{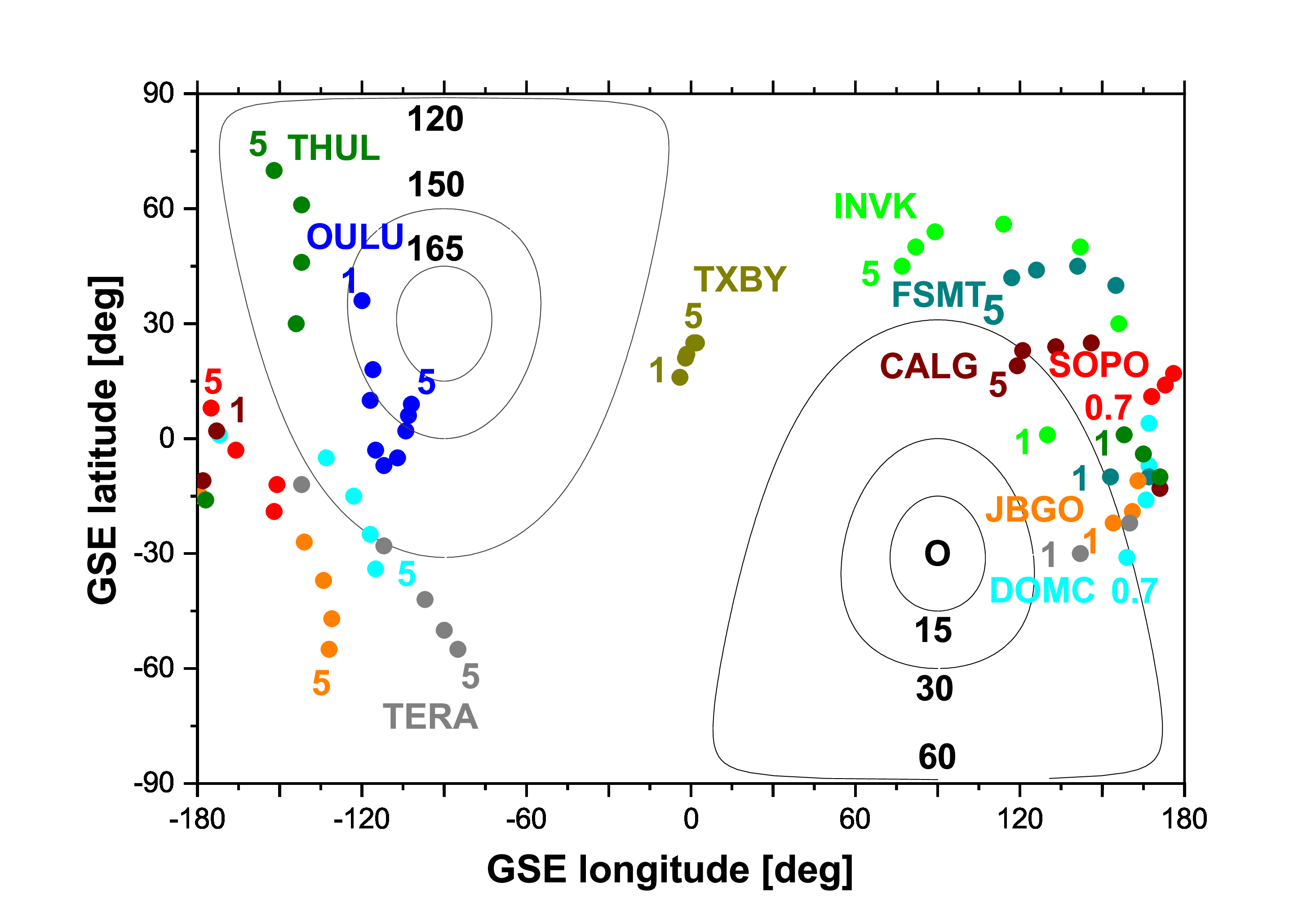}
\caption{Asymptotic directions in GSE coordinates for selected NMs stations during GLE74. The asymptotic directions are plotted along with the NM acronyms in the rigidity range $\sim$1--5 GV. The lines of equal pitch angles relative to the derived anisotropy axis are plotted for 15$^{\circ}$, 30$^{\circ}$, 60$^{\circ}$ for sunward directions, and 165$^{\circ}$, 150$^{\circ}$ and 120$^{\circ}$for anti-Sun direction.} 
\label{fig:Cones}
\end{figure}

In Figure \ref{fig:Cones} we present an illustration of the asymptotic directions computed for selected NM stations in the rigidity range of 1--5 GV. This range corresponds to the maximal NM response, whereas in the analysis we used the 1--20 GV rigidity range. 

\subsection{Modelling the response of neutron monitors}  
Using the new-generation NM yield function calibrated using PAMELA \citep{Adriani2017} and AMS-02 \citep[Alpha Magnetic Spectrometer --][]{Aguilar2021c} records, details given by \citet{koldobsky19, Koldobskiy20222585}, and explicitly considering the station scaling factor, we modelled the global NM network response using the expression: 

\begin{equation}
\frac{\Delta N}{N}(P_{\mathrm{cut}},t) = \frac{\int_{P_{\mathrm{cut}}}^{P_{\mathrm{max}}}J_{\mathrm{SEP}}(P,t)S(P)G(\alpha,t) A(P)dP}{\sum_{i}\int_{P_{\mathrm{cut}}}^{\infty}J_{\mathrm{GCR}_{i}}(P,t)S_{i}(P)dP}
\label{eq:nmresponse}
   \end{equation}
where $\Delta N$ is the NM count rate increase produced by SEPs, $N$ is the NM count rate, that is the background, produced by GCR, $J_{\mathrm{SEP}}$ is the rigidity spectrum of SEPs, accordingly $J _{\mathrm{GCR}_{i}}(P,t)$ is the rigidity spectrum of the $i$ component (proton or $\alpha$-particle, etc...) of GCR at given time $t$. $G(\alpha,t)$ is the pitch angle distribution (PAD), that is $sic.$ "the distribution along the angle between the axis of symmetry of the particle distribution and the asymptotic viewing direction at rigidity $P$, associated with the arrival direction" \citep{Cramp97, Bombardieri2007813}. $A(P)$ is a discrete function with $A(P)$=1 and 0 for allowed and forbidden trajectories, respectively \citep{Cook91}. $P_{\mathrm{cut}}$ is the minimum rigidity cut-off of the station, $P_{\mathrm{max}}$ is the maximum rigidity of SEPs considered in the model (20 GV), whilst for GCR $P_{\mathrm{\mathrm{max}}}$= $\infty$. $S$ is the NM yield function. The GCRs are represented by the force-field model \citep{Cab04,usoskin2005heliospheric}, considering all species, using the local interstellar spectrum parameterised by \citet{Vos2015}, and the modulation potential according to \citet{UsoskinJGR2017}. 

\begin{figure}[h!]
\centering
\includegraphics[width=\columnwidth]{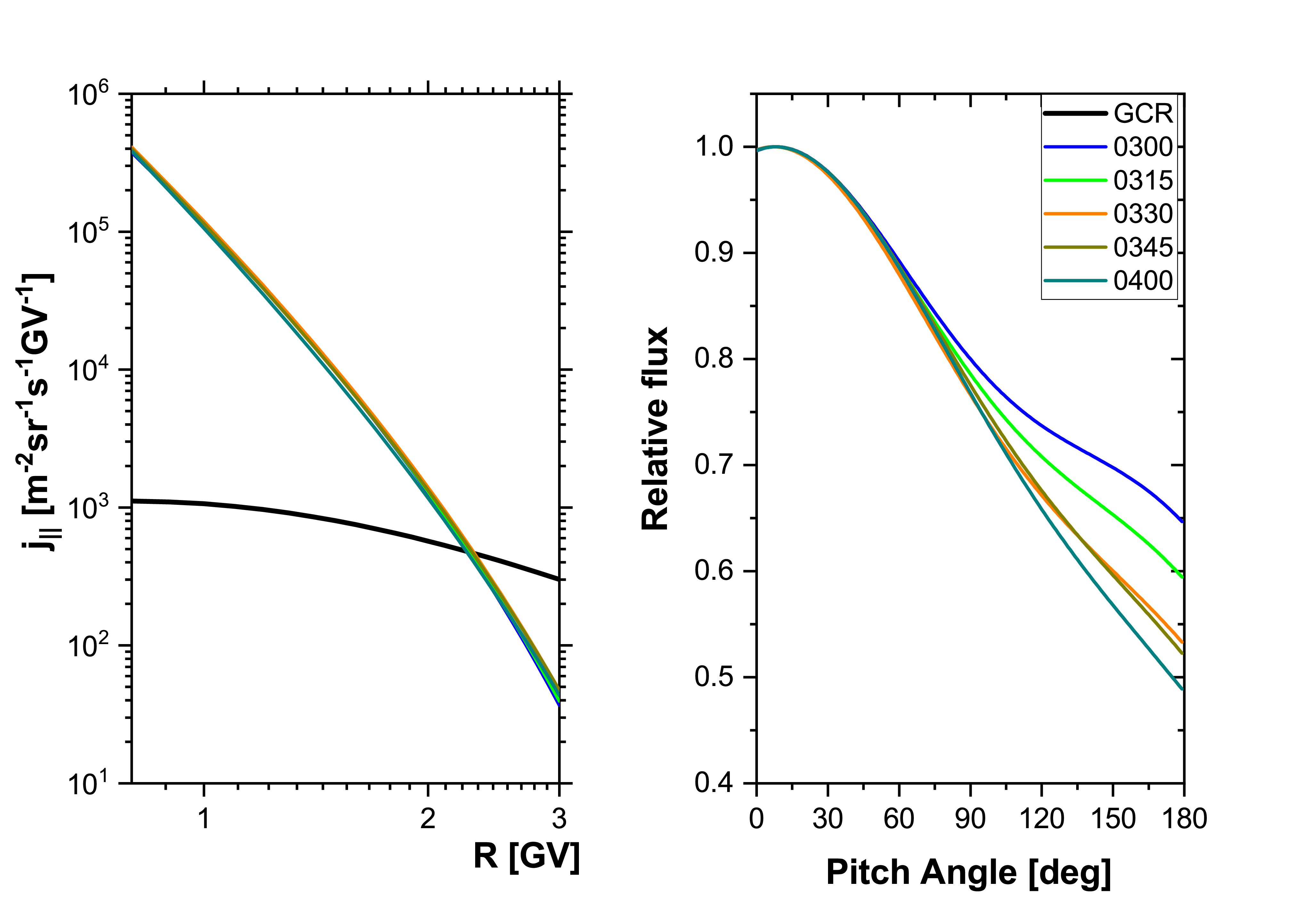}
\caption{Reconstructed spectra and PAD of SEPs during various stages of the GLE.} 
\label{fig:specpad}
\end{figure}

We note that the event under study in this work bears some resemblance to the second Halloween event, specifically GLE66 on October 29, 2003, in terms of its occurrence during a deep FD and a severe geomagnetic storm. Therefore, herein we explicitly considered the FD and the major geomagnetic storm effect on the rigidity cut-off, similarly to \citet{Larsen2025}. According to our analysis, the best fit of the SEP spectral distributions is achieved by using a modified power-law function:
\begin{equation}
j_\parallel(P) = j_{0}P^{-(\gamma+\delta\gamma(P-1{\,\rm GV}))},
\label{Eq:sp1}
\end{equation}
\noindent where $j_\parallel(P)$ is the differential particle flux parallel to the axis of symmetry, where $j_0$ is the differential SEP intensity at $P$ = 1 GV.  The $\gamma$ and $\delta\gamma$ are the power-law index and its steepening, respectively. Accordingly, the best fit for the PAD is achieved with double Gaussian, that is, the incoming SEPs from both the Sun and anti-Sun directions are given by:
\begin{equation}
        G(\alpha) \approx \exp(-\alpha^{2}/\sigma_{1}^{2}) + B \cdot \exp(-(\alpha-\pi)^{2}/\sigma_{2}^{2}), 
        \label{Eq:pad}
   \end{equation}
where $\alpha$ is the pitch angle, $\sigma_{1}$ and $\sigma_{2}$ are parameters that determine the width of the distributions. The term $B$ accounts for the contribution of particles arriving from the anti-Sun direction. An illustration of several spectra and PAD reconstructed throughout the event are presented in Fig. \ref{fig:specpad}. 

The SEP spectra were moderately hard with slopes $\gamma$ ranging from about 5 during the event onset to about 6.3 at the late phase of the event. A significant roll-off of the spectra $\delta\gamma$, that is steepening in the high-rigidity/energy part, was observed, which gradually diminished throughout the event, but never vanished. A notable anti-Sun SEP flux was observed, which resulted in a relatively broad, not a beam-like, angular distribution, specifically during the main phase (maximum particle flux) stage of the event, namely the $B$ parameter of about 0.5, $\sigma_{1}$ ranging from 3.2 during the event onset to about 6.5 in the late phase of event, and $\sigma_{2}$ from 2.5 to about 5.9 for the initial and late phase, respectively. 

\section{Near-Earth measurements of GLE74} \label{sec:nearEarth}

\begin{figure}[h!]
\centering
\includegraphics[width=0.87\columnwidth]{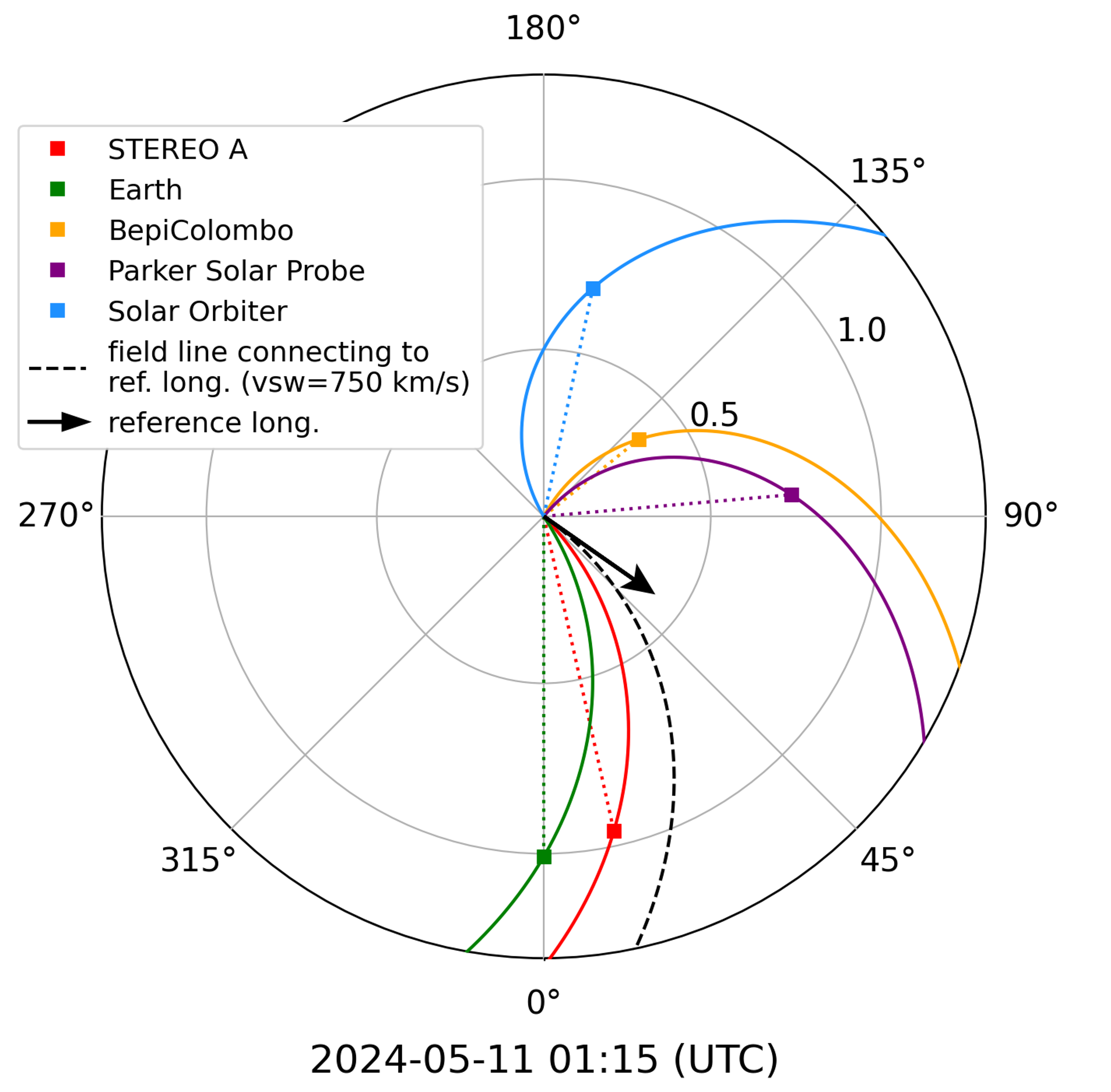}
\caption{View of the heliographic equatorial plane from solar north, showing the positions of various spacecraft on 11 May 2024 at 01:15~UT. The Parker spirals are shown for each spacecraft. The data are from the Solar MAgnetic Connection Haus tool \citep[\url{https://solar-mach.github.io},][]{2023FrASS...958810G}.} 
\label{fig:sc_pos}
\end{figure}

High energy SEPs were clearly recorded by particle instruments on near-Earth orbiting spacecraft, including EPHIN and ERNE \citep[Energetic and Relativistic Nuclei and Electron][]{torsti1995energetic} on board SOHO, GOES/SEISS, and the High Energy Telescope (HET) of STEREO-A \citep{2008SSRv..136..391V}. Figure~\ref{fig:sc_pos} shows the positions of various spacecraft in the heliosphere and the Parker spirals connecting each location, at the time of GLE74 ($\sim$01:15~UT). A solar wind speed of 790~km/s measured by the Advanced Composition Explorer \citep[ACE,][]{Stone1998} at L1 and $\sim$720 km/s from STEREO/PLASTIC was used for Earth and STEREO-A, respectively. Using these measured solar wind speeds, we calculated the location of the footpoints of the nominal Parker spirals for Earth and STEREO-A. The footpoints connected to Earth and STEREO-A were located  at W32 and W46, respectively (in the HGS system at 01:15~UT).

\begin{figure}[h!]
\centering
\includegraphics[width=\columnwidth]{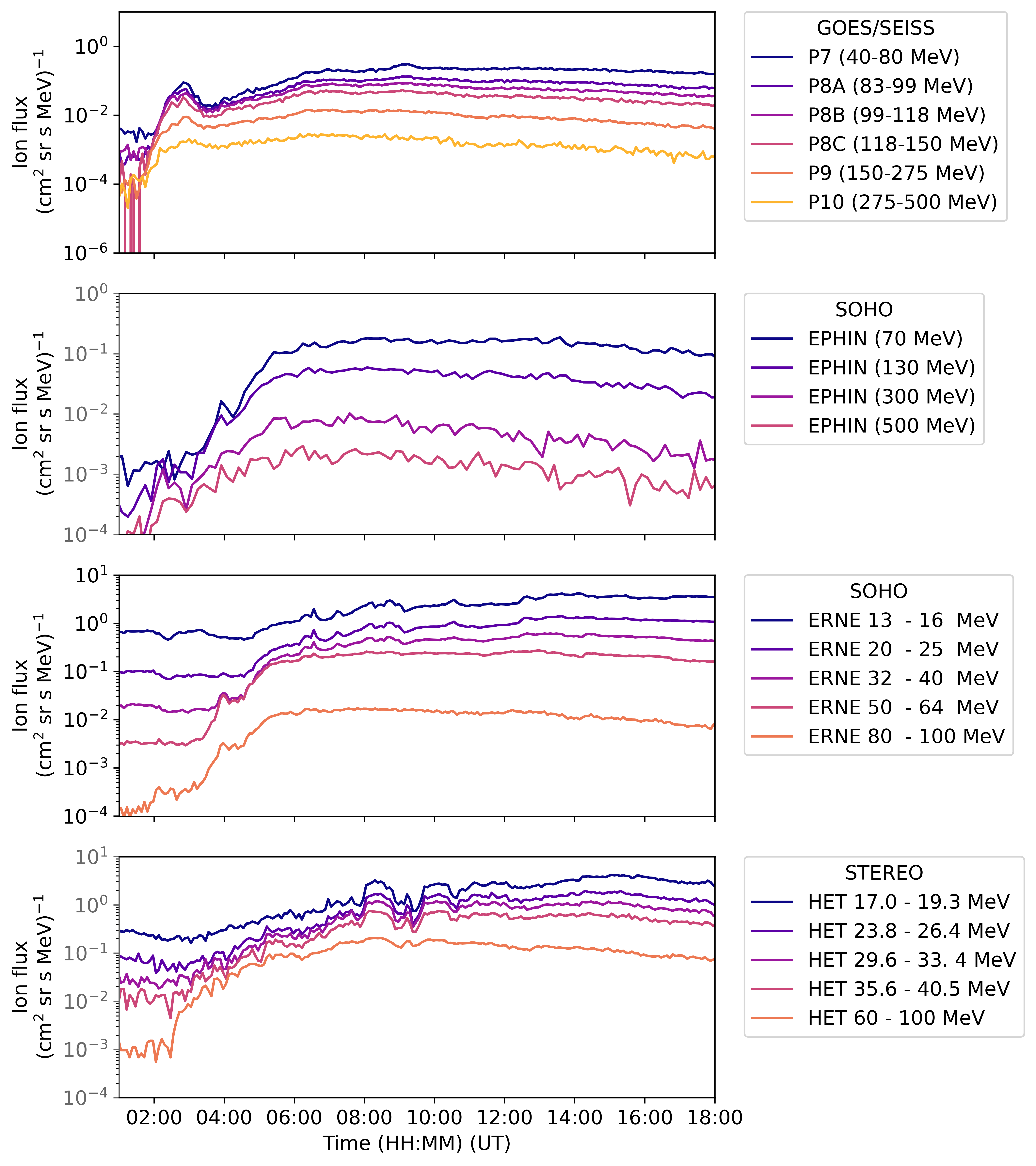}
\caption{Energetic particle recordings of GLE74 in the near-Earth space. From top to bottom: GOES/SEISS (40--500 MeV), SOHO/EPHIN (70--500 MeV), SOHO/ERNE (13--100 MeV), and STEREO-A/HET (17--100 MeV) measurements, respectively.}  
\label{fig:GOEs}
\end{figure}

The analysis of GLE74 in this work is focused on the near-Earth spacecraft and STEREO-A, which is the least separated from Earth (see Fig.~\ref{fig:sc_pos}). The time history of SEP measurements during GLE74 as recorded from GOES/SEISS (40--500 MeV), SOHO/EPHIN (70--500 MeV), SOHO/ERNE (13--100 MeV), and STEREO-A/HET (26.3--100 MeV) are presented in Fig.~\ref{fig:GOEs}. High-energy protons at each spacecraft seem to have a prompt increase: GOES/P10 (275--500 MeV) has an onset time at 01:15~UT$\pm$5min and SOHO/EPHIN (at 500 MeV) records the event at 01:24~UT$\pm$10min.

\section{Relation to solar sources}
For the first arriving particles, it is possible to perform a time-shifting analysis \citep[TSA,][]{vainio2013first} to infer their release time at the Sun, i.e., the solar release time (SRT). A low-end energy limit of particles recorded by a sea-level NM station is $\sim$1~GV (i.e., 433 MeV), and thus the corresponding mean velocity for such energetic protons would be $u = 0.73c$. For GLE74, particles with rigidities of up to $\sim$2.4~GV (1.6~GeV) have been identified, with a mean velocity of $u = 0.93c$. 

The length of the Parker spiral, $L$, can be computed based on the solar wind speed during the event \citep{vainio2013first, paassilta2017catalogue}. During GLE74, the solar wind speed was fast, $V_\mathrm{SW}$=790~km/s, leading to $L$ = 1.04 AU. For the first arriving particles, we assumed scatter-free propagation and calculated the expected SRT of the relativistic protons, $t_\mathrm{rel}$, adding 500~s for comparison with remote-sensing measurements at 1~AU \citep[e.g., radio observations;][] {2022A&A...660L...5P}. 

Due to the fact that GLE74 evolved on the background of a very disturbed period, the onset time of the GLE on the ground by NMs may be ambiguous. As a result, very high energy particles from SOHO/EPHIN at E=500 MeV were used instead of NM recordings for the TSA. The earlier registered onset was obtained at $t_\mathrm{onset}$ = 01:24~UT (see Table~\ref{tab:timeline}). The travel time of these protons ($v=0.75c$) was calculated to be $\sim$11.4 min and the corresponding anticipated  $t_\mathrm{rel}$ is $\sim$01:21 UT. For comparison, the relativistic protons of $\sim$2.4~GV ($v=0.93c$) have a travel time of $\sim$9.3 min for the same Parker spiral.

\begin{figure}[h!]
\centering
\includegraphics[width=\columnwidth]{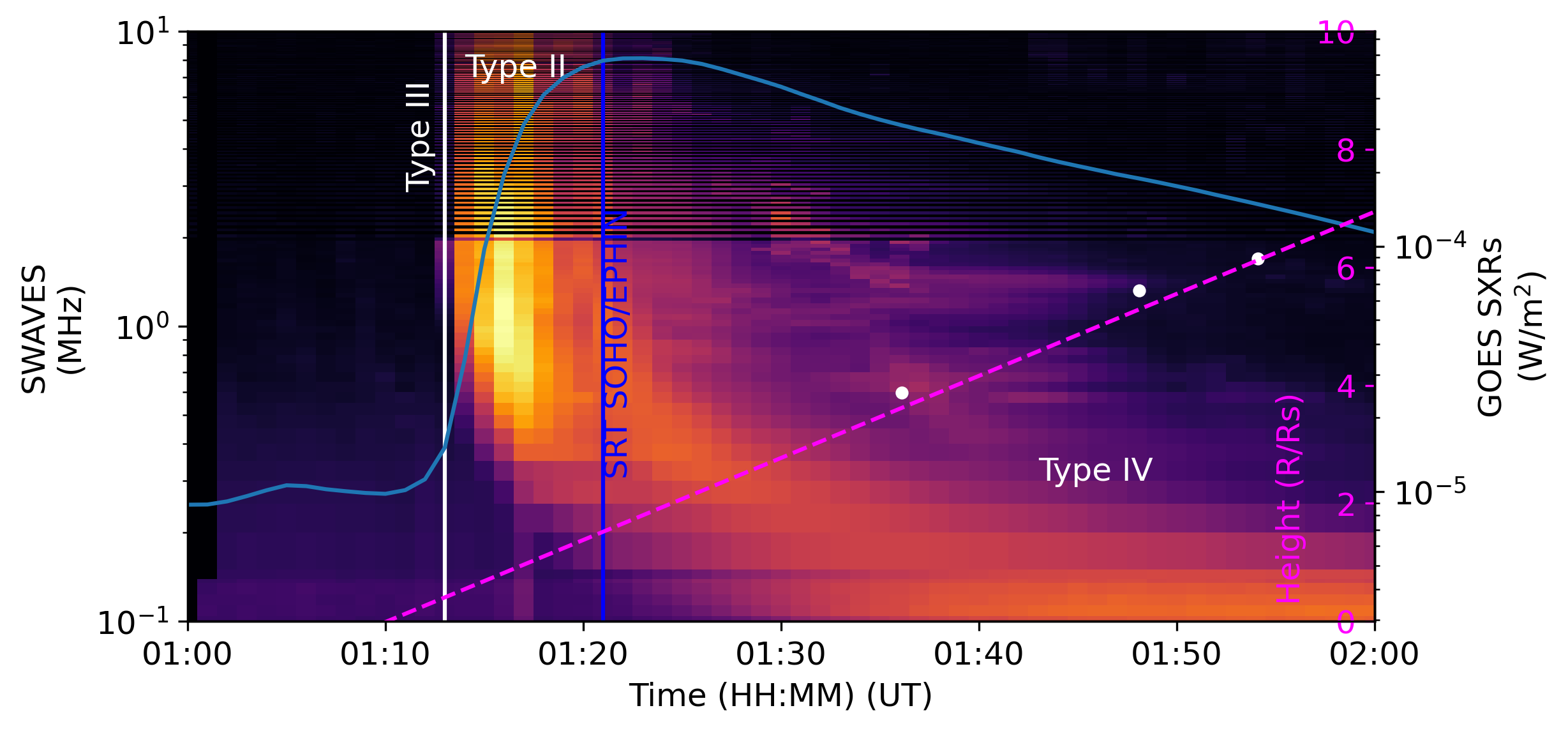}
\caption{A radio spectrogram from SWAVES. On top of which are overplotted: the GOES SXRs denoting the X5.8 flare (light blue solid line, y-axis on the right-hand external side); the first of the series type III burst is indicated by a vertical white line. The type II burst and the interplanetary type IV are further indicated on the plot. Data from the height-time measurements from SOHO/LASCO are imprinted as white circles and the linear fit to the points is presented with magenta color (y-axis on the right-hand internal side). The blue vertical line indicates the upper limit SRT based on SOHO/EPHIN.} 
\label{fig:solar}
\end{figure}

From STEREO-A/WAVES (SWAVES) data\footnote{\url{https://stereo.space.umn.edu/data/level-3/STEREO/Ahead/SWAVES/one-minute/IDL/HFR-LFR/2024/}} and ground-based observatories from e-Callisto\footnote{\url{https://www.e-callisto.org/}} we find that there is a group of type III bursts at the initiation of the event. The start time of the type III radio burst is marked at 01:12~UT (see Fig. \ref{fig:fig1}(d)), while the start of a type II radio burst is set to 01:13~UT\footnote{\url{https://soleil.i4ds.ch/solarradio/data/BurstLists/2010-yyyy_Monstein/2024/e-CALLISTO_2024_05.txt}}. The coronal type II radio burst begins at a frequency of approximately 350 MHz. It extends into the interplanetary medium around $\sim$400 kHz. The measured drift rate is $\sim$0.5 MHz/s (from the ALMATY station). A metric type IV radio burst was recorded by e-Callisto stations, such as ALMATY, and extended into the decametric-hectometric range, as detected by Wind/WAVES and STEREO/SWAVES between 01:42 and 03:55 UT.

Comparing with the soft X-ray (SXR) and radio observations, we find that the release of $E=$500 MeV particles ($\sim$01:21~UT) occurred $\sim$2~minutes prior to the flare peak time, 9~minutes after the start of the first type III radio burst and $\sim$8~min after the type II radio burst onset (Fig.~\ref{fig:fig1} and Fig.~\ref{fig:solar}). Around the release time of the energetic protons there is radio emission from a group of the type III radio bursts and a moving type IV radio burst that appears latter on (see Fig.~\ref{fig:solar}). At the release time of the $E=$500 MeV particles, the CME is located at a height of $\sim$1.8~$R_{sun}$ (see Fig.~\ref{fig:solar}). 
Table \ref{tab:timeline} provides a timeline of events during GLE74 based on the measurements and subsequent calculations. The flare onset at SXRs (01:10~UT) closely aligns with the Type III burst (01:12~UT), indicating an initial burst of electron acceleration. The Type II burst (01:13~UT) suggests that the CME-driven shock formed quickly, propagating outward with a drift rate of 0.5 MHz/s. The appearance of the CME at 01:36:05 UT (3.88 $R_{sun}$) in LASCO C2 supports a fast shock formation and expansion, consistent with a high-speed CME (1614 km/s). The Type IV burst (01:25 UT – 03:55 UT) reflects sustained electron trapping and emissions within the CME-driven shock, transitioning into the interplanetary medium.

\begin{table}[!ht]
    \centering
\caption[]{Timeline of events for GLE74. }
\label{tab:timeline}

    \begin{tabular}{l l}
    
    \hline
        Event & Time [UT] \\ 
        \hline
        \hline
        SXR onset & 01:10 (1min)\\ 
        Type III onset (first of the group) & 01:12 (1sec) \\
        Type II onset & 01:13 (1sec) \\
        GOES/SEISS onset ($E$=275--500 MeV) & 01:15 (5min) \\
        SRT ($E$=500 MeV) & 01:21 (10min)\\
        SXR peak & 01:23 (1min)\\
        SOHO/EPHIN onset ($E$=500 MeV) & 01:24 (10min) \\
        Type IV (Metric) &  01:25 (1min) \\
        GLE onset at South Pole & 01:35 (5min)\\
        CME first observation in LASCO/C2 & 01:36 ($\sim$5min) \\
        \hline
        \multicolumn{2}{p{0.9\columnwidth}}{\textbf{Notes}. All times are Earth times, and propagation times for electromagnetic emissions have been considered in this table as explained in the text. The numbers in parentheses denote the time resolution of the measurements used.}
    \end{tabular} \label{tab:time}
\end{table}


\subsection{Comparison with other GLEs}
\label{sec:NMcom}
\begin{figure}[h!]
\centering
\includegraphics[width=\columnwidth]{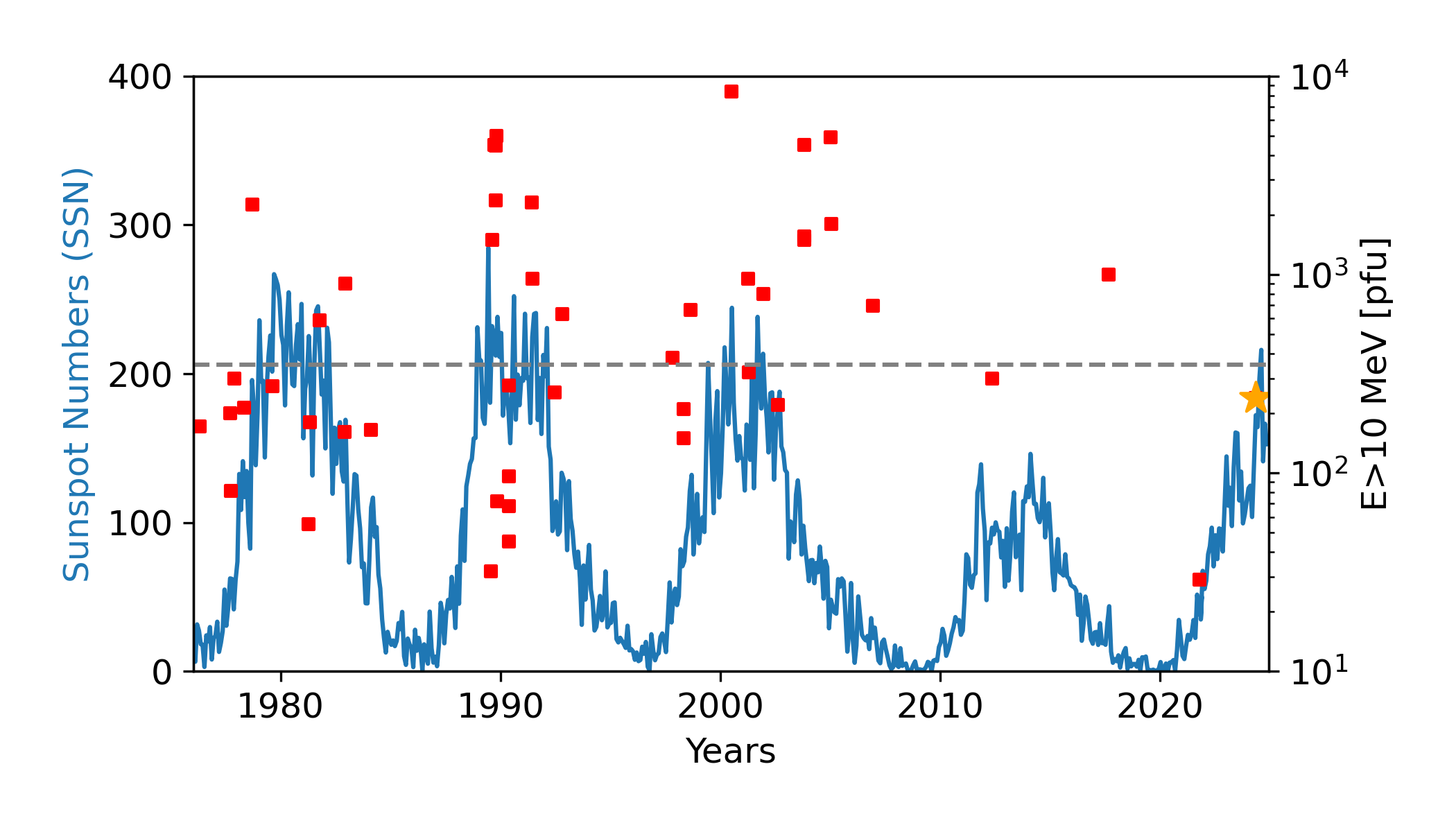}
\caption{Peak proton flux ($I_{P}$) at E$>$10 MeV for GLEs that occurred from 1976 to 2024 (red squares). GLE74 is denoted by an orange star. The median (350 pfu) of the peak proton flux is presented as a horizontal dashed line. Monthly sunspot numbers are shown as a blue line.} 
\label{fig:sc}
\end{figure}

\begin{figure*}[h!]
\centering
\includegraphics[width=0.49\columnwidth]{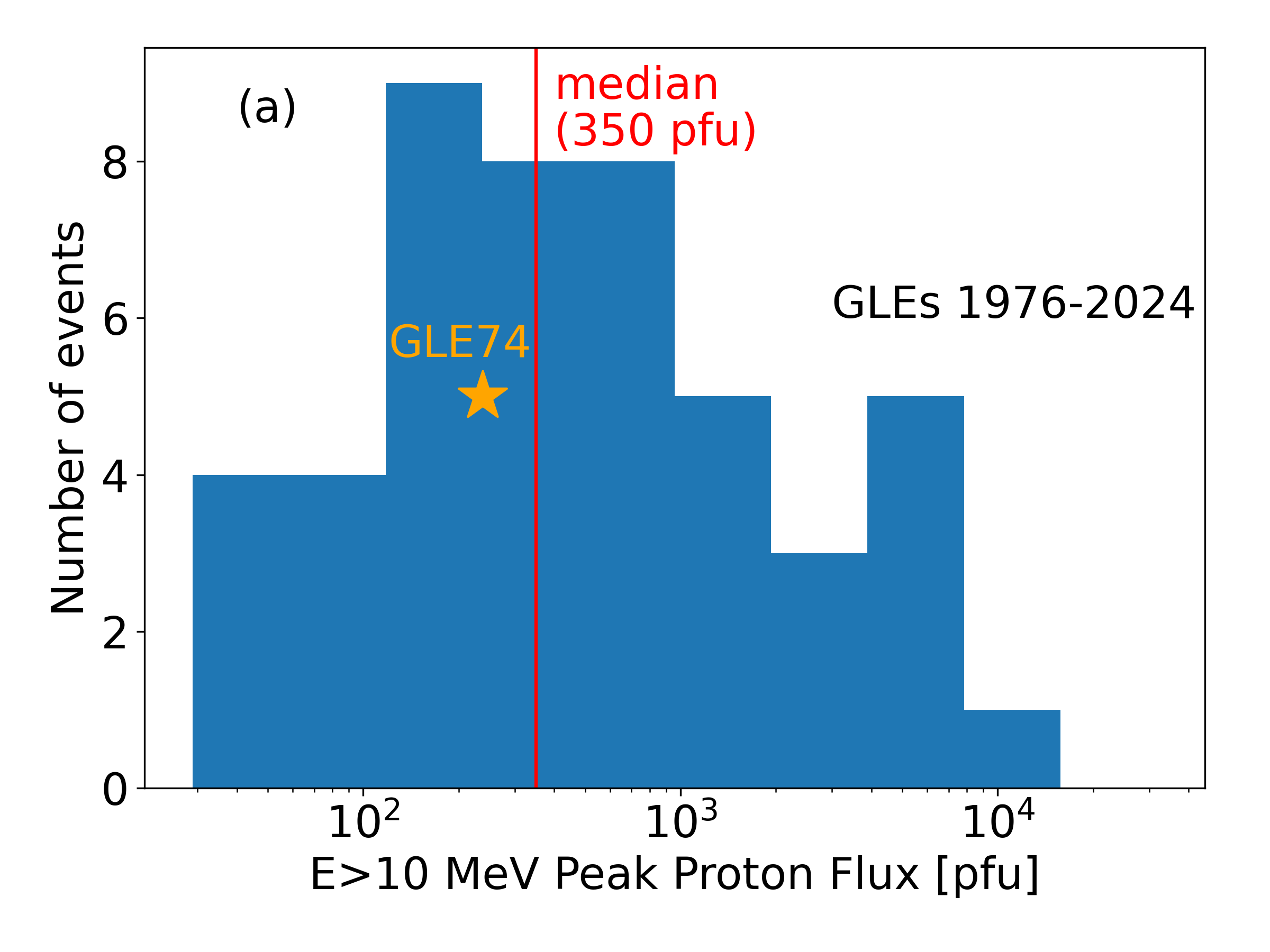}
\includegraphics[width=0.49\columnwidth]{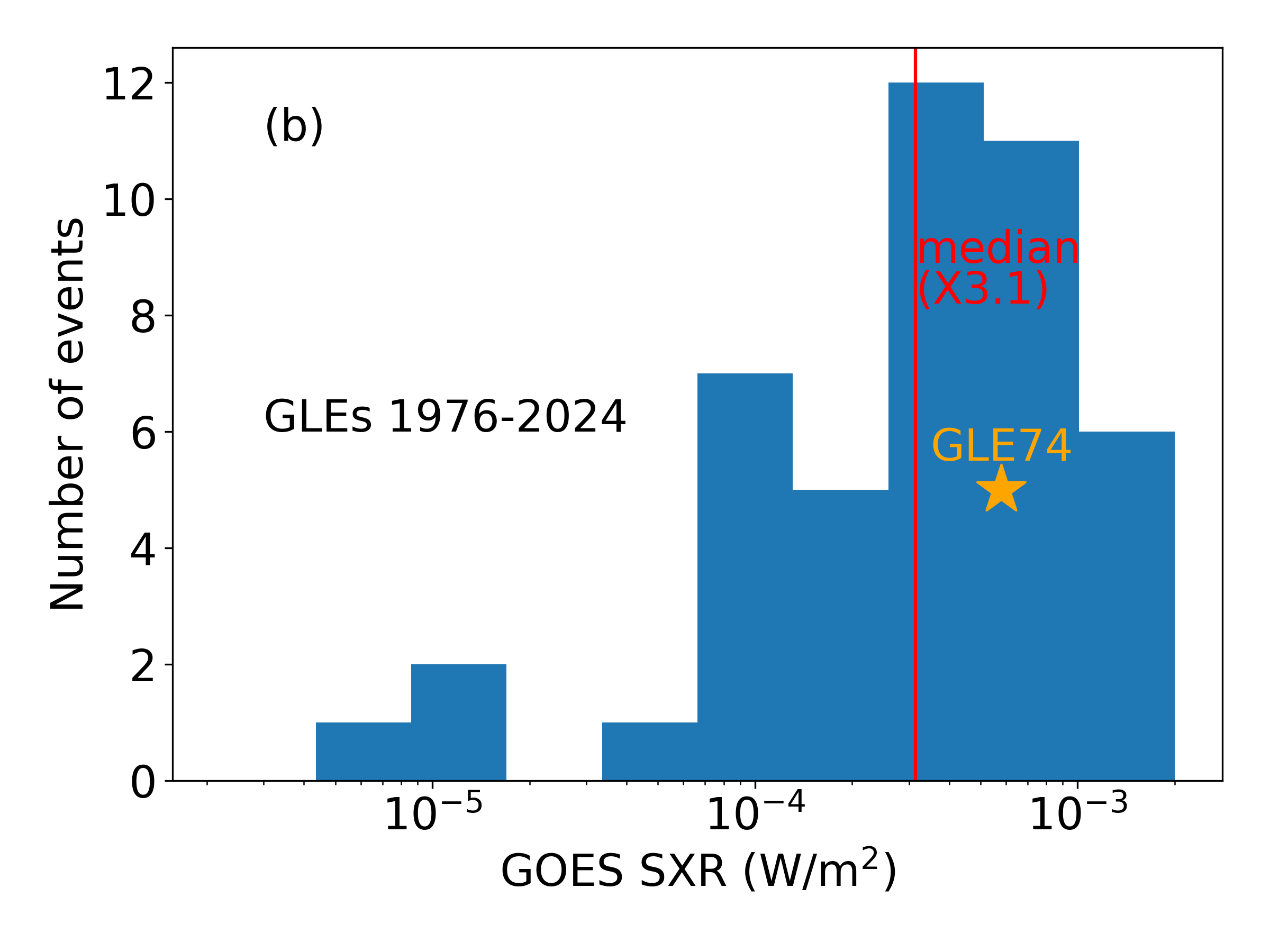}
\includegraphics[width=0.49\columnwidth]{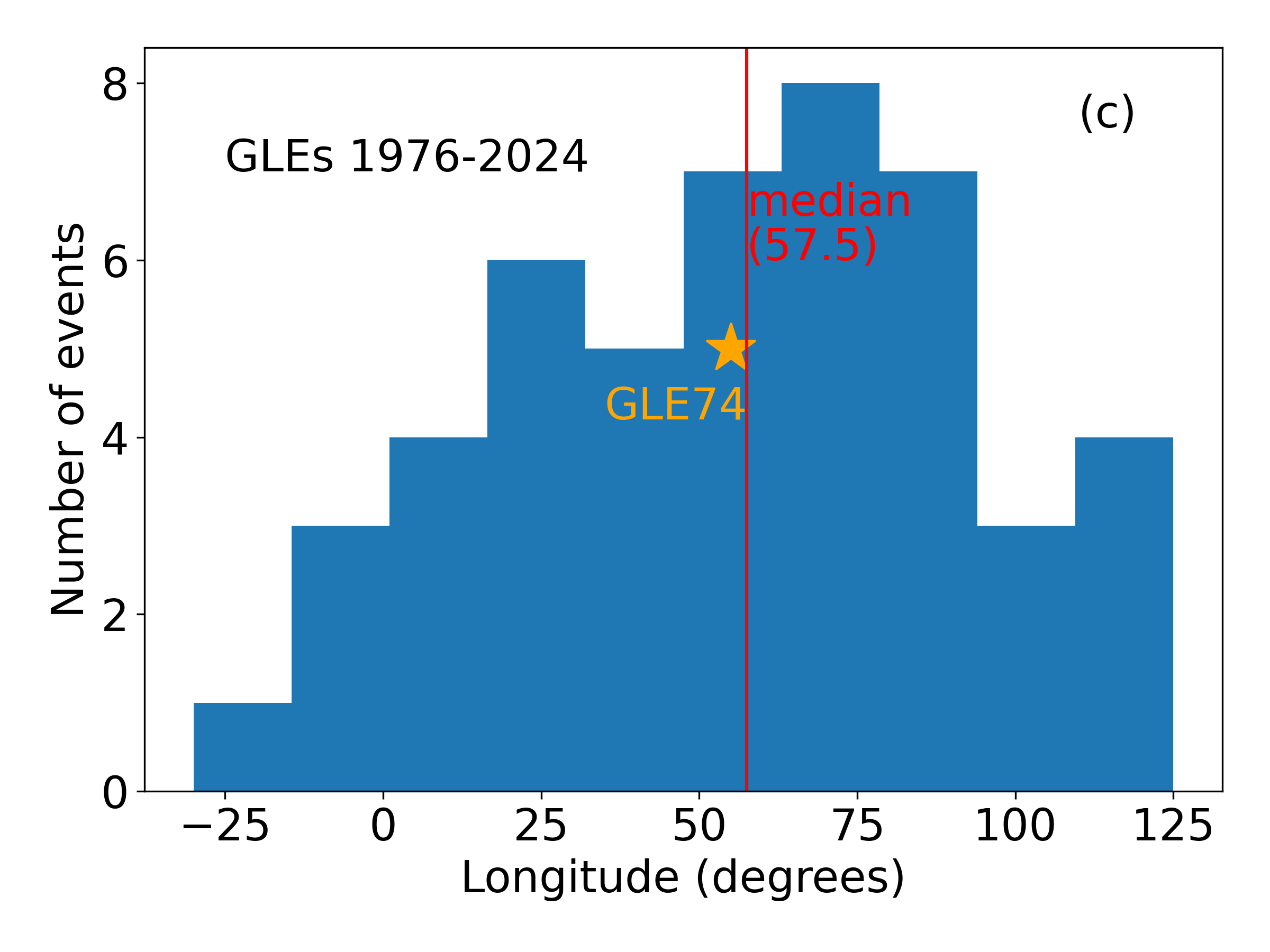}
\includegraphics[width=0.49\columnwidth]{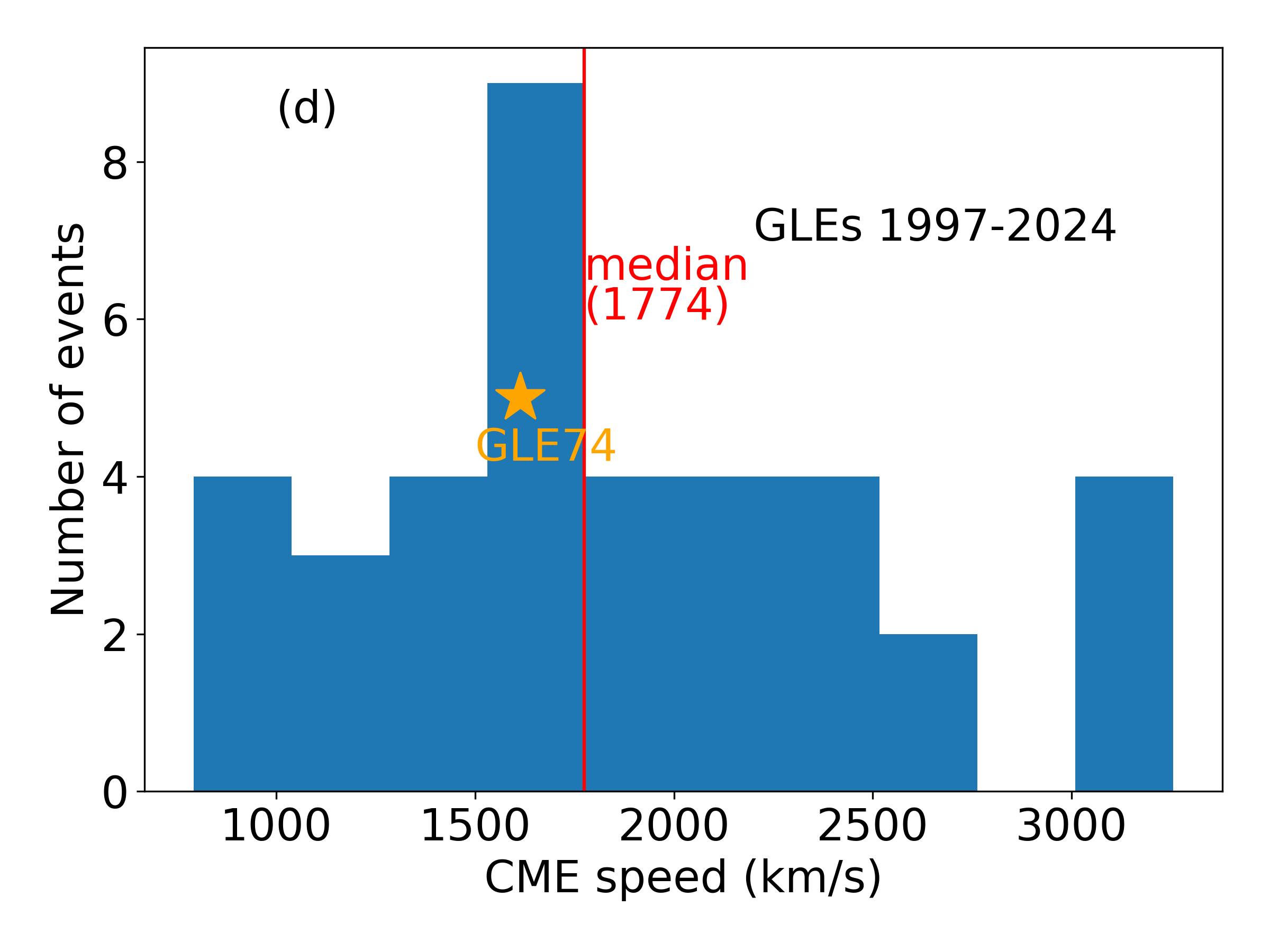}
\caption{Distribution of observables related to GLEs that occurred from 1976 to 2024. GLE74 is denoted in all plots by an orange star. The median value of each parameter is presented in each panel as a vertical red solid line and it is further imprinted on the plot: (a) the peak proton flux ($I_{P}$) at $E>$10 MeV (in pfu); (b) GOES SXR (W/m$^{2}$); longitude of the associated flare (in degrees) and (d) CME speed (in km/s) .} 
\label{fig:stat}
\end{figure*}

Since 1976 and up until May 2024, a total of 47 GLE events have been recorded \footnote{\url{https://gle.oulu.fi/}}. Figure \ref{fig:sc} shows the time distribution from 1976--2024 of the peak flux of SEPs with $E$>10 MeV ($I_{P}$), as detected by GOES\footnote{\url{https://www.ngdc.noaa.gov/stp/space-weather/interplanetary-data/solar-proton-events/SEP\%20page\%20code.html}}, for the GLE events is plotted with red squares, on the background of the evolving solar cycle (blue trace) as marked by the sunspot numbers (SSN\footnote{Source: WDC-SILSO, Royal Observatory of Belgium, Brussels [\url{https://www.sidc.be/SILSO/datafiles}]}). The median and mean values of the GLE $E$>10 MeV peak proton fluxes are found to be 350 pfu and 1840 pfu, respectively, indicating a distribution skewed toward higher flux events. GLE74, indicated in the figure with an orange star, had a peak proton flux of $I_{P} = 238$ pfu, which is closely aligned with the median value (represented by the dashed gray vertical line), suggesting it was a moderately intense event in this category. Moreover, Figure \ref{fig:stat} provides a detailed statistical analysis of the GLEs since 1976, focusing on several key parameters. Panel (a) provides a distribution of $I_{P}$, demonstrating results similar to Fig. \ref{fig:sc}. Panel (b) presents the distribution of GLEs in relation to their associated SXR flux. GLE74 was associated with an X5.8-class solar flare, which is near the median SXR flux of X3.1 (depicted as a red vertical line in this panel). This further supports the classification of GLE74 as a typical event, based on its flare association.
Furthermore, panel (c) illustrates the longitudinal distribution of the associated solar flares (in degrees), an important factor influencing the efficiency of particle acceleration and transport \citep{klein2017acceleration}. GLE74 originated from a W55 (in degrees) solar longitude, which is well within the median range of previously observed GLEs (W57.5, red vertical line in panel (c)). Finally, panel (d) displays the CME speed distribution, measured in km/s, which serves as another key indicator of solar eruptive activity \citep{2012SSRv..171...23G}. The CME linked to GLE74 had a speed of 1614 km/s, which is close to the median value of the entire dataset, reinforcing its classification as a relatively standard event in terms of solar eruption dynamics.

\begin{table}[!ht]
    \centering
\caption[]{GLE74 compared to median values of GLEs}
    \begin{tabular}{l c c c c}
    \hline
        Event& Longitude  & SXR  & CME speed  & $I_{P}$ \\ 
             & [degrees] & [W/m$^{2}$] & [km/s] & [pfu] \\
        \hline
        \hline
        GLE74 & 55 & X5.8 & 1614 & 238\\
        Median & 57.5 & X3.1 & 1530 & 350\\
        \hline
    \end{tabular} \label{tab:GLEs}
\end{table}
Table \ref{tab:GLEs} provides a summary of the median values of the key parameters investigated in this study, along with the corresponding values for GLE74. As evident from the data, GLE74 aligns well with these median values across multiple characteristics, indicating that it exhibits behavior typical of past GLEs. These findings are consistent with previous studies that have examined the statistical properties of GLE events and their correlations with solar eruptive phenomena \citep{2012SSRv..171...23G, macau_mods_00083703}.

\section{Conclusions}

In this work a summary of observations for GLE74 -- which took place on 11 May 2024 -- is presented. Detailed modelling and reconstruction of the spectral and angular characteristics of high-energy SEPs in the vicinity of the Earth were performed. Data from ground-based NMs, together with space-borne data, were employed in the corresponding data analysis. Additionally, the increase in NM count rates during GLE74 was influenced by a complex interplay between the direct signal from solar particles, the recovery of the FD, and the complex geomagnetospheric conditions. The main results of the study are:

\begin{enumerate}
    \item During the main phase of GLE74, the rigidity spectrum exhibit moderate hardness, with slopes ($\gamma$) ranging from approximately 5 to $\sim$6.3. 
    
    \item A notable spectral rollover ($\delta\gamma$), characterized by a steepening in the high-rigidity/energy region, was observed. This steepening gradually weakened over time but never completely disappeared.
    
    \item  A notable SEP flux from anti-Sun direction was detected, exhibiting a relatively broad angular distribution—rather than a narrow, beam-like pattern—particularly during the main phase of the event, when particle flux reached its peak.
    
    \item The SRT of the very high-energy particles onboard SOHO (EPHIN; $E$=500 MeV) was found to be $\sim$01:21~UT, and around this SRT the CME-driven shock was located at a height of $\sim$1.8 ($\pm0.2$)~$R_{sun}$.

    \item A series of type III bursts (starting at 01:12~UT), a type II (onset at 01:13~UT) and a type IV (onset at SWAVES at 01:42~UT) burst were identified in conjunction to GLE74.
    
\end{enumerate}
Based on the observational evidence presented in this study, a key finding is the observed anti-Sun flux detected by NMs and the anisotropy seen in the intensity time profiles across different NMs (see Fig.~\ref{fig:figaniso}). This is suggestive of certain magnetic field configurations that may have contributed to these observed fluxes. Specifically, departing from the standard Archimedean/Parker spiral, potential candidates include a magnetic bottleneck beyond Earth and a closed interplanetary magnetic loop. In the latter case, particles could be injected along the near leg of the loop and reflected in the far leg or injected along both legs of the loop \citep[see details in ][and references therein]{2006ApJ...639.1186R}. The presence of multiple CMEs and their interplanetary counterparts prior to and during GLE74 is further suggestive of such possibilities. Further investigation through detailed transport and 3D CME-shock modeling is necessary to accurately determine the underlying scenario \citep[see e.g.][]{2016ApJ...833...45R}.

The SRT of the near-relativistic particles at 01:21~UT($\pm$10min) is in agreement with the actual SXR peak flux time (01:23~UT$\pm$1min), indicating a delay between the energetic (rising) phase of the flare (see Table \ref{tab:timeline} and Fig.~\ref{fig:solar}) \citep[similar cases for  GLEs are reported in][]{2002ApJ...567..622B, 2004ApJ...601L.103B,2006ApJ...639.1186R}. Additionally, near the SRT the CME-driven shock was located at a height of $\sim$1.8 ($\pm0.2$)~$R_{sun}$. Hence, it can be suggested that the release of relativistic particles could be attributed to acceleration at a CME shock that requires time to form \citep{2012SSRv..171...23G}. Nonetheless, such an interpretation further requires the examination of composition data \citep{2024A&A...682A.106K}, coupled with finer time resolution. 
%
\begin{acks}
This research received funding from the European Union’s Horizon Europe programme under grant agreement No 101135044 (SPEARHEAD) [\url{https://spearhead-he.eu/}].Views and opinions expressed are however those of the author(s) only and do not necessarily reflect those of the European Union or the European Health and Digital Executive Agency (HaDEA). Neither the European Union nor the granting authority can be held responsible for them.

The present work benefited from discussions held at the International Space Science Institute (ISSI, Bern, Switzerland) within the frame of the international team: gRound and spacE-bAsed analySis of Strong sEp eventS and Study of their terrestrial effects (REASSESS) [\url{https://teams.issibern.ch/reassess/}]. This study was partly supported by the Research Council of Finland project No. 354280 GERACLIS and the Horizon Europe Program project ALBATROS and the National Science Fund of Bulgaria under contract KP-06-H64/3. M.L. acknowledges the Space It Up project funded by the Italian Space Agency, ASI, and the Ministry of University and Research, MUR, under contract n. 2024-5-E.0 - CUP n. I53D24000060005.

We acknowledge the NMDB database (\url{www.nmdb.eu}) founded under the European Union's FP7 programme (contract no. 213007), and the PIs of individual neutron monitors. Italian polar program PNRA (via the LTCPAA PNRA 2015/AC3 and the BSRN PNRA OSS-06 projects), the French Polar Institute IPEV and FINNARP are acknowledged for the hosting of DOMB/DOMC NMs. 

A.P.R, N.T and M.J. also acknowledge support from the French space agency (Centre National des Etudes Spatiales; CNES), the plasma physics data center (Centre de Données de la Physique des Plasmas; CDPP; \url{http://cdpp.irap.omp.eu/} and the Solar-Terrestrial Observations and Modelling Service; STORMS; \url{ http://storms-service.irap.omp.eu/ }.

\end{acks}

\bibliographystyle{spr-mp-sola}
\bibliography{fd.bib}  

\end{article}
\end{document}